\documentclass[aps, prd, twocolumn,tightenlines, notitlepage, superscriptaddress, nofootinbib, preprintnumbers, floatfix,showkeys,10pt,altaffilletter]{revtex4-2}

\usepackage{afterpage}

\usepackage{upgreek}
\usepackage{placeins}

\usepackage[official]{eurosym}
\usepackage[normalem]{ulem}
\usepackage{amstext}
\usepackage{graphicx}
\graphicspath{{plots/}}
\usepackage{url}
\usepackage{color}
\usepackage{ulem}
\usepackage[version=4]{mhchem}
\usepackage[utf8]{inputenc}
\usepackage{fontawesome}
\usepackage{yfonts}
\usepackage{amsmath}

\pdfoutput=1
\usepackage{textcomp}
\usepackage{comment}
\usepackage{yfonts}
\usepackage{epsfig,amsfonts,mathrsfs,graphicx,color,slashed,multirow}
\usepackage{latexsym,graphicx,slashed,color,enumerate,url,cancel,gensymb}
\usepackage{textcomp}

\usepackage{slashed}
\usepackage{aas_macros}

\usepackage[dvipsnames]{xcolor}
\usepackage[colorlinks,pdfstartview=FitV,breaklinks=true]{hyperref}
\usepackage{booktabs}
\usepackage{adjustbox}
\usepackage{textgreek} 
\usepackage{caption, subcaption} 
\usepackage{ragged2e} 
\DeclareCaptionJustification{justified}{\justifying}

\usepackage{booktabs} 
\usepackage{float}
\usepackage{orcidlink}
\usepackage{lmodern}
\usepackage{ae,aecompl}
\usepackage{appendix}
\usepackage{accents}

\usepackage{tikz}

\usepackage[capitalise]{cleveref}

\definecolor{mediumpersianblue}{rgb}{0.0, 0.4, 0.65}

\AtBeginDocument{\hypersetup{citecolor=mediumpersianblue,linkcolor=mediumpersianblue,urlcolor=mediumpersianblue}}
\usepackage{appendix}

\begin{document}

\newcommand{\diff}{\ensuremath{\mathrm{d}}}

\newcommand{\Ledd}{\ensuremath{L_\mathrm{Edd}}}
\newcommand{\Msun}{\ensuremath{M_\odot}}
\newcommand{\Mdot}{\ensuremath{\dot{M}}}
\newcommand{\MPBH}
{\ensuremath{M_\mathrm{PBH}}}
\newcommand{\MBH}
{\ensuremath{M_\mathrm{BH}}}
\newcommand{\MH}
{\ensuremath{M_\mathrm{h}}}
\newcommand{\fPBH}
{\ensuremath{f_\mathrm{PBH}}}

\newcommand{\GeV}
{\ensuremath{\mathrm{GeV}}}
\newcommand{\cm}
{\ensuremath{\mathrm{cm}}}
\newcommand{\gr}
{\ensuremath{\mathrm{g}}}
\newcommand{\pc}
{\ensuremath{\mathrm{pc}}}
\newcommand{\Rsun}
{\ensuremath{R_\odot}}

\newcommand{\rsp}
{\ensuremath{r_\mathrm{sp}}}
\newcommand{\rhosp}
{\ensuremath{\rho_\mathrm{sp}}}
\newcommand{\rhoDM}
{\ensuremath{\rho_\mathrm{DM}}}
\newcommand{\rhob}
{\ensuremath{\rho_\mathrm{bkg}}}
\newcommand{\rhoSun}
{\ensuremath{\rho_\odot}}


\preprint{}

\title{Hints of Dark Matter Spikes in Low-mass X-ray Binaries:\\a critical assessment}

\author{Francesca Scarcella}
\email{scarcella@ifca.unican.es}

\author{Bradley J. Kavanagh}
\email{kavanagh@ifca.unican.es}

\affiliation{\IFCA}

\newcommand{\IFCA}{Instituto de F\'isica de Cantabria (IFCA, UC-CSIC), Av.~de
Los Castros s/n, 39005 Santander, Spain}


\begin{abstract}

Three black-hole low-mass X-ray binaries (LMXBs) in the Milky Way show rates of period decay which cannot be easily explained by standard mechanisms. 
Recently, it has been claimed that the anomalous period decays in two of these systems may be explained by dynamical friction due to very high dark matter (DM) densities around the black holes.  We critically assess these claims by performing $N$-body simulations of binaries embedded in dense DM ``spikes". We simulate the previously-studied systems XTE J1118+480 and A0620--00, as well as studying the third binary Nova Muscae 1991 for the first time in this context. These simulations show that feedback on the DM distribution plays a crucial role and we rule out previously-claimed shallow DM spikes. We set lower limits on the steepness $\gamma$ of DM density profiles required to explain the period decay in these LMXBs, requiring $\gamma \gtrsim 2.15-2.20$ in XTE J1118+480 and A0620--00 and  $\gamma \gtrsim 2.3$ in Nova Muscae 1991. Improved modeling of the long-term evolution of binaries embedded in DM spikes may allow us to exclude even larger densities in future. 

\end{abstract}

\maketitle

\section{Introduction}

Dark Matter (DM) remains one of the great mysteries of modern particle physics, astrophysics and cosmology~\cite{Bertone:2004pz}.  The environments around Black Holes (BHs) offer a promising arena for searching for the influence of DM, and therefore for understanding its nature~\cite{Bertone:2018krk,Baryakhtar:2022hbu}. This is due in part to the possibility that the DM density around BHs may be many orders of magnitude larger than, for example, in the diffuse halos of galaxies. If BHs grow adiabatically at the centres of dense DM halos, they may grow a power-law `spike' of DM with a density profile growing as $r^{-7/3}$~\cite{Gondolo:1999ef,Ullio:2001fb,Sadeghian:2013laa} (or a somewhat flatter `mound', depending on the exact process of formation and growth~\cite{Bertone:2024wbn}). It has also been proposed that BHs may be formed from the collapse of large overdensities in the early Universe -- so-called Primordial Black Holes (PBHs)~\cite{Zeldovich:1967lct,Hawking:1971ei,Green:2020jor} -- and spikes are expected around all such objects, with a steeper power-law index of $9/4$~\cite{Mack:2006gz,Eroshenko:2016yve,Adamek:2019gns,Carr:2020mqm}. In these environments, DM can reach densities as high as $\mathcal{O}\left(10^{19}\right)\,M_\odot \,\mathrm{pc}^{-3} \sim\mathcal{O}\left(10^{-3}\right)\,\mathrm{g} \,\mathrm{cm}^{-3}$~\cite{Gondolo:1999ef,Bertone:2024wbn}, over 20 orders of magnitude larger than the density of DM in the Solar neighbourhood~\cite{deSalas:2020hbh}. 

Such large overdensities of DM can have a number of observational consequences. If DM consists of self-annihilating particles, these dense spikes would give rise to observable fluxes of Standard Model particles, for example, gamma rays~\cite{Lacki:2010zf,Bertone:2019vsk}. DM spikes can also enhance the rate of accretion of BHs which they surround~\cite{Mack:2006gz,Ricotti:2007au,DeLuca:2020fpg,Serpico:2020ehh,Jangra:2024sif}. Here, we focus on the purely gravitational influence that DM spikes may have on the dynamics of binary systems. The dominant effect is typically expected to be dynamical friction (DF)~\cite{Chandrasekhar:1943I,Chandrasekhar:1943II,Chandrasekhar:1943III}, in which the gravitational deflection of DM particles around the orbiting object leads to a drag force which can accelerate the inspiral. This effect is potentially observable as a modification of the phase evolution of Gravitational Waves (GWs) from compact object binaries (see e.g.~\cite{Macedo:2013qea,Eda:2013gg,Eda:2014kra,Barausse:2014tra,Barausse:2014pra,Cardoso:2019rou,Coogan:2021uqv,Becker:2022wlo,Cole:2022yzw,Zhang:2024ugv,Karydas:2024fcn,Vicente:2025gsg}) or as a modification of the merger rates of such binaries~\cite{Kavanagh:2018ggo,Yue:2018vtk,Jangra:2023mqp}.\footnote{Similar effects due to clouds of ultra-light bosons around black holes have also been studied in e.g.~Refs.~\cite{Baumann:2021fkf,Baumann:2022pkl,Tomaselli:2023ysb,Tomaselli:2024dbw}.} Instead, for nearby binaries which are detected in electromagnetic radiation, the accelerating effect of DM spikes on the inspiral could be directly probed by measurements of the orbital period (and its derivative). 

Indeed, a number of black hole low-mass X-ray Binaries (BH LMXBs)~\cite{Remillard:2006fc} have been observed with anomalously fast orbital decays. These are XTE J1118+480~\cite{GonzalezHernandez:2011} (hereafter System A), A0620--00~\cite{Hernandez:2013} (hereafter System B) and Nova Muscae 1991~\cite{2017MNRAS.465L..15G} (hereafter System C). Each of these systems consists of a star ($M_\star \sim 0.1 - 0.9\,M_\odot$) tightly orbiting a stellar mass BH ($M_\mathrm{BH} \sim 6 - 11\,M_\odot$), with the accretion of material from the star by the BH leading to observable X-ray emission. Crucially, the period of these three systems is decreasing  more than 100 times faster than expected from the emission of GWs alone, while there is no consensus that other mechanisms, such as magnetic braking or mass loss, can explain this discrepancy. 

Chan \& Lee have interpreted the anomalous binary evolution of XTE J1118+480 and A0620--00 in terms of dynamical friction from DM, suggesting the presence of spikes in these systems with power-law indices $\gamma = 1.85$ and $\gamma = 1.71$ respectively~\cite{Chan:2022gqd}.\footnote{The impact of DM spikes on the orbital decay of SMBH binaries has also been studied in, e.g., Refs.~\cite{Alachkar:2022qdt,2024ApJ...962L..40C,Deb:2025raq}.} Subsequently, Ref.~\cite{Kar:2023vqe} explored the consequences of these proposed spikes for signals from DM annihilation. However, these studies neglected the effects of feedback on the evolution of the DM spike; the spike may be substantially depleted by the motion of the binary, reducing the effects of dynamical friction. 
Ref.~\cite{Ireland:2024lye}, which considered the possibility that the BHs in XTE J1118+480 and A0620--00 may be primordial in origin, took into account feedback using a semi-analytic prescription presented in Ref.~\cite{Kavanagh:2020cfn}. However, subsequent simulations have suggested that this prescription may substantially over-estimate the amount of feedback in such systems~\cite{Mukherjee:2023lzn,Kavanagh:2024lgq}.

In this work we aim to critically evaluate a Dark Matter interpretation of the anomalous period decays using \textsc{NbodyIMRI}~\cite{NbodyIMRI,Kavanagh:2024lgq}, a dedicated $N$-body code designed for studying the dynamics of binaries embedded in DM spikes. The output of these simulations allows us to infer the evolution of the period due to dynamical friction induced by the DM while accounting for feedback effects on the DM spike. In turn, this allows us to infer the initial DM density profile which best matches the observed orbital decay, once depletion of the spike is taken into account. In addition to XTE J1118+480 and A0620--00, we also study Nova Muscae 1991 for the first time in this context. This is especially important because the component masses of Nova Muscae 1991 are more similar than in the other two systems, which is expected to lead to larger feedback effects. Finally, we discuss whether the densities inferred from our simulations are expected in common spike formation scenarios in the literature and explore how much of a role DM could play in these anomalous observations.

In \cref{sec:Anomalies}, we describe in more detail the systems which we study in this paper and discuss braking  mechanisms, including dynamical friction.
In \cref{sec:Spike} we discuss the spike modelling and in \cref{sec:Setup} we summarise the \textsc{NbodyIMRI} framework which we use to study the evolution of the three systems.
We present our simulation results in \cref{sec:Results}, including the inferred period evolution and DM density profiles for these systems. Finally, in \cref{sec:Discussion}, we critically evaluate a Dark Matter interpretation of these results,
before concluding in \cref{sec:Conclusions}.

\section{Anomalous period decay of galactic X-ray binaries}
\label{sec:Anomalies}

\subsection{Observations}

 A few hundred X-ray binaries have been identified in the Milky Way, in $\sim$ twenty of which the compact object has been confirmed to be a BH~\cite{Fortin:2024siz}. Most of these are low mass X-ray binaries (LMXB), systems in which the donor is a small ($\lesssim 1 \,\Msun$) star.
 Accurate measurements of the orbital period of BH LMXBs can be obtained through the Doppler variation of the stellar spectrum. 
 Measuring variations in the orbital period is challenging and requires repeated high-precision (spectral) observations over a long time baseline. Observations spanning 15 -- 20 years have allowed to measure the period decay rate of three BH-LMXB systems~\cite{GonzalezHernandez:2011,Hernandez:2013,2017MNRAS.465L..15G}; 
 their properties are listed in~\cref{tab:observations}. The three systems present similar characteristics: a light star tightly orbiting a BH, with a period of a few hours, decaying at a rate of $\mathcal{O}(1 - 10)\,\mathrm{ms/yr}$. In all three cases, the period-decay rate has been found to be anomalously high, compared to the predictions from standard orbital decay mechanisms.

\begin{table*}[th!]
\centering
\begin{tabular}{cccc}
    \hline\hline &&&\\[-1.2em]
     &  \textcolor{OliveGreen}{\textbf{System A}}   &  \textcolor{RoyalBlue}{\textbf{System B}}   &  \textcolor{Sepia}{\textbf{System C}}   \\  & XTE J1118+480 & A0620-00 & NM1991 \\
    \hline &&&\\[-1em]
    $M_{\rm BH}$ [$\Msun$] & $7.46^{+0.34}_{-0.69}$ \cite{Hernandez:2013} & $5.86\pm0.24$ \cite{vanGrunsven:2017nua} & $11.0^{+2.1}_{-1.4} \cite{2017MNRAS.465L..15G}$ \\
    \,\,\,$M_\star$ [$\Msun$]\,\,\, & $0.18\pm0.06$ \cite{Hernandez:2013} &  $0.40\pm0.01$ \cite{Hernandez:2013}  & $0.89\pm0.18$ \cite{2017MNRAS.465L..15G} \\
    $r_\star$ [$R_\odot$] & $0.34\pm0.05$ \cite{2017MNRAS.465L..15G} & $0.67\pm0.02$\cite{2017MNRAS.465L..15G} & $1.06\pm0.07$ \cite{2017MNRAS.465L..15G}  \\
    \,\,\,$q = M_*/M_{\rm BH}$\,\,\, & $0.024 \pm 0.009$ \cite{Hernandez:2013} & $0.060 \pm 0.004$ \cite{vanGrunsven:2017nua}  & $0.079\pm0.007$ \cite{2017MNRAS.465L..15G} \\[0.2em]
    \hline &&&\\[-1.2em]
    $r_{\rm orb}$ [$R_\odot$] & $2.54\pm0.06$ & $3.79\pm0.04$ & $5.49\pm0.32$\\
    $P_0$ [day]  & $0.16993404(5)$ \cite{Hernandez:2013} & $0.32301415(7)$ \cite{Hernandez:2013} &  $0.432605(1)$ \cite{2017MNRAS.465L..15G}  \\
    $\dot{P}$ [ms/yr]  & $-1.90 \pm 0.57$ \cite{Hernandez:2013} &  $-0.60 \pm 0.08$ \cite{Hernandez:2013} & $-20.7\pm12.7~\cite{2017MNRAS.465L..15G} $ \\[0.2em]
    \hline\hline 
\end{tabular}
\caption{\textbf{Observational constraints on the LMXBs} discussed in this work. We report one-sigma constraints on the BH mass $M_\mathrm{BH}$, stellar mass $M_\star$, stellar radius $r_\star$, mass ratio $q$, and orbital radius $r_\mathrm{orb}$. $P_0$ is the first measurement of the orbital period and $\dot{P}$ is the period time derivative. }
\label{tab:observations}
\end{table*}

\subsection{Braking mechanisms}

The standard evolution of short-period LMXBs attributes angular momentum loss to gravitational radiation, systemic mass loss (ML), and magnetic braking (MB)~\cite{1981A&A...100L...7V, 1982ApJ...253..908T, 1982ApJ...254..616R}. Mass which is lost from the system (ML) reduces the angular momentum of the binary at fixed semi-major axis, leading to a shrinking of the binary and a decay of the orbital period. Meanwhile, MB arises because the ionised material of the stellar wind is confined to flow outwards along the magnetic field lines of the companion star. Because of this, the stellar wind corotates with the star, leading to a loss of angular momentum when the material escapes at large distances.

For systems with orbital periods of a few hours, gravitational radiation contributes at the level $|\dot{P}| \lesssim 0.01 \, \mathrm{ms}/\mathrm{yr}$; MB and ML combined are expected to produce $|\dot{P}|$ of order $\lesssim 0.02 \, \mathrm{ms}/\mathrm{yr}$, according to standard prescriptions used in binary evolution studies~\cite{GonzalezHernandez:2011}.
Observed values for XTE J1118+480 and A0620–00 and the much larger -- albeit less certain -- value reported for Nova Muscae 1991 (see Table~\ref{tab:observations})
thus exceed standard channels by one to two orders of magnitude.
Proposals to explain the anomalous orbital‑period decays invoke enhanced or convection/rotation‑boosted magnetic‑braking prescriptions~\cite{GonzalezHernandez:2011,2024ApJ...976..210F} and resonant tidal torques from circumbinary disks~\cite{2015A&A...583A.108C, 2018ApJ...859...46X, 2019ApJ...876L..11C}. Combinations of these mechanisms can plausibly reproduce the measured $\dot{P}$ in XTE J1118+480 and A0620–00 under specific parameter choices. However, the most extreme value reported for Nova Muscae 1991 remains difficult to explain and no consensus has yet emerged on a model accounting for all observations.

In this context, some authors have recently interpreted these anomalies as signatures of dark‑matter spikes around the black holes, attributing the orbital‑period decay to dynamical friction.

\subsection{Dynamical friction}

A star (or, in general, a massive body) moving through a background of lighter particles experiences a drag force of purely gravitational origin. This force, known as dynamical friction, arises as the cumulative effect of a large number of weak gravitational interactions between the star and the particles.
Since, in the star's rest frame, the velocity distribution of the latter appears skewed (particles travel \textit{towards} the star) momenta are exchanged preferentially along the direction of the orbital velocity. This exchange gives rise to a force opposing the stellar motion.
Dynamical friction was first described by Chandrasekhar~\cite{Chandrasekhar:1943I,Chandrasekhar:1943II,Chandrasekhar:1943III} in the context of the relaxation of stellar systems. He obtained his foundational formula by computing the star's momentum variation per unit time due to two-body hyperbolic encounters, integrated over impact parameters and relative velocities. Chandrasekhar's expression for dynamical friction reads
\begin{equation}
\label{eq:DF}
     \mathcal{F}_\mathrm{DF} = - \dfrac{4 \pi G_\mathrm{N}^2 \, M_\star^2 \, \rho \, \xi(v) \, \ln \Lambda}{v^2} \, ,
\end{equation}
where $\rho$ is the density of background particles; $M_\star$ and $v$ are the mass and velocity of the star; $\xi(v)$ is the fraction of particles moving slower than the orbital speed $v$; and $\Lambda$ is the ratio of the maximum and minimum impact parameters. Relativistic extensions of Eq.~\eqref{eq:DF} have also been studied (see e.g.~\cite{Chiari:2022kas,Vicente:2025gsg}). 

As a consequence of dynamical friction, the star loses part of its kinetic energy and is obliged to move to less demanding orbits, sinking toward the centre of the gravitational potential. In the case of a star on a circular orbit, the gradual shrinking of the orbital radius $r_\mathrm{orb}$ is accompanied by an increase of the angular velocity, hence a reduction of the orbital period:
\begin{align}
\label{eq:Pdot}
    \begin{split}
        \dot{P} &= v \mathcal{F}_\mathrm{DF} \frac{\mathrm{d}P}{\mathrm{d}E}= -6 G_\mathrm{N}\frac{M_\star}{M_\mathrm{BH}}P^2 \rho(r_\mathrm{orb}) \, \xi(v)\,  \ln \Lambda \,,
    \end{split}
\end{align}
where $P$ is the orbital period and $\left|E\right| = G_\mathrm{N}M_\mathrm{BH}M_\star/(2r_\mathrm{orb})$ is the orbital energy.
This physical mechanism has been invoked in previous studies~\cite{Chan:2022gqd,Kar:2023vqe,Ireland:2024lye} to attribute the anomalous period decay of systems A and B to the presence of a DM spike around the BH: that is, a DM ``dress'' with a steep density profile $\rho_\mathrm{DM} \propto r^{-\gamma}$. In these works, the value of $\rho_\mathrm{DM}$ at the orbital radius was inferred from the observed decay rate through~\cref{eq:DF} and used to constrain $\gamma$.

However,~\cref{eq:DF} does not capture the effects of feedback on the DM spike~\cite{Kavanagh:2020cfn,Mukherjee:2023lzn,Karydas:2024fcn,Kavanagh:2024lgq}. The energy which is extracted from the binary by dynamical friction is injected into the DM spike, altering the orbits of the DM particles and depleting the DM density close to the orbiting object. This in turn can dramatically reduce the size of the dynamical friction effect. 

\begin{figure}[tb]
\centering
\hspace{-0.5em}
\includegraphics[width=0.97\linewidth]{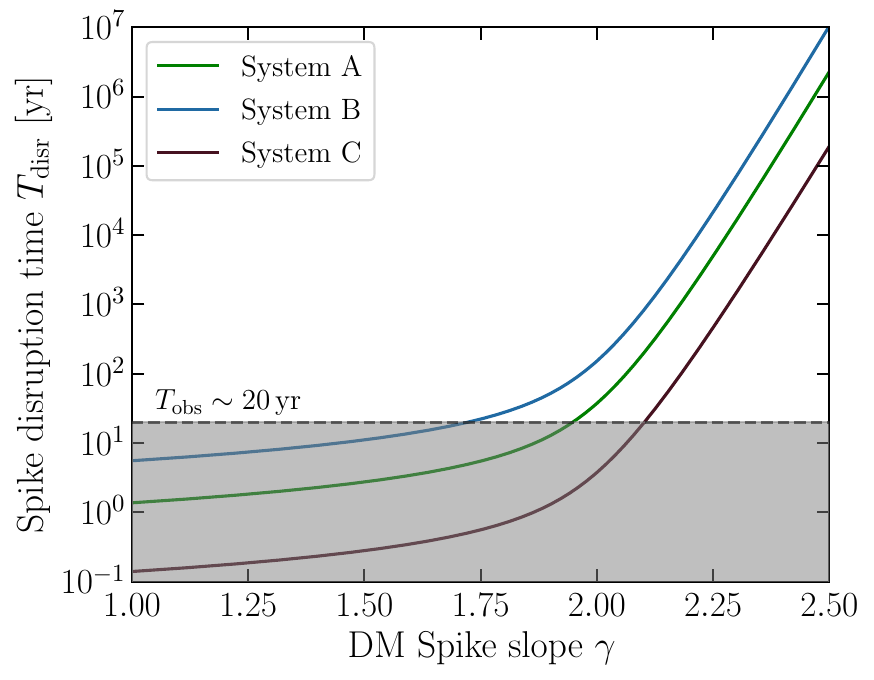}
\caption{\textbf{Estimated disruption timescales for DM spikes in the three binary systems.} This timescale is obtained by comparing the total gravitational binding energy of a spike with slope $\gamma$ to the rate of energy injection inferred from the observed value of $\dot{P}$. A DM-related explanation of the anomalous inspiral rates is ruled out in the grey shaded region; here, the spike is likely to be completely disrupted on timescales shorter than the duration of observation of these systems, $T_\mathrm{obs} \sim 20 \,\mathrm{yr}$.}
\label{fig:disruption_timescale}
\end{figure}

As a first estimate of the importance of these effects, we calculate the total gravitational binding energy of the DM spike (following the prescription of Ref.~\cite{Kavanagh:2020cfn}, with full details of the assumed density profile in Sec.~\ref{sec:Simulations}). This binding energy represents the reservoir of available work that can be done by dynamical friction to speed up the inspiral of the binary. If the observed value of $\dot{P}$ is entirely due to DF, we can estimate the rate of energy injection into the spike as $\dot{E} = (\mathrm{d}E/\mathrm{d}P)\times\dot{P} = (2E/3P)\times \dot{P}$. Assuming a constant value of $\dot{P}$, we can estimate the disruption timescale $T_\mathrm{disr}$: the time taken for the work done by DF to exceed the available binding energy of the spike, leading to its complete disruption. 

Numerical estimates of this disruption timescale as a function of the spike slope $\gamma$ are shown in Fig.~\ref{fig:disruption_timescale}. For large values of $\gamma$, the density of DM close to the BH is enhanced, leading to a larger binding energy and disruption timescales on the order of $T_\mathrm{disr} \sim 10^5-10^7\,\mathrm{yr}$ for $\gamma = 2.5$. However, for shallower spikes ($\gamma \lesssim 2$), the disruption timescale can be shorter than the time over which observations of these binary systems have been recorded (grey shaded region). For this range of $\gamma$ (which includes values inferred in previous studies~\cite{Chan:2022gqd,Kar:2023vqe}), the DM spike does not contain sufficient energy to support the anomalously large values of $\dot{P}$ observed in these systems over the past 15-20 years.


It is clear from this preliminary calculation that testing any DM interpretation of the anomalous period decays requires taking into account how energy injection affects the DM distribution. In practice, the depletion of the DM spike reduces the DF force and thus the rate of energy injection, a feedback process that our first simple estimate does not account for.
These feedback effects have previously been studied in the context of intermediate- and extreme-mass-ratio inspirals (IMRIs/EMRIs), but are expected to become even more pronounced for larger mass ratios $q \gtrsim 10^{-2}$, as seen in the systems we study here. We will therefore rely on $N$-body simulations to study the spike response and the resulting impact on $\dot{P}$.

\section{Simulations of binary evolution within dark matter 
spikes}
\label{sec:Simulations}

\subsection{Spike model and initialisation}
\label{sec:Spike}

The DM spike density is assumed to grow towards the BH following a power-law profile, starting from a radius $\rsp$.
Following the customary prescription~\cite{Eda:2014kra,2004PhRvL..92t1304M}, we estimate the extension of the spike $\rsp$  as $0.2$ times the radius of gravitational influence $r_\mathrm{in}$, where the latter is defined by the condition
\begin{equation}
M_\mathrm{DM}( \leq r_\mathrm{in})=\int_0^{r_\mathrm{in}}4 \pi r^2 \rhoDM \left(r \right) \diff r=2 \MBH \, ,
\end{equation}
where $\rhoDM$ is the DM density, $\MBH$ is the black hole mass.
By assuming that the background density $\rhob$ is recovered at the spike radius $\rsp$, the normalisation of the spike is fixed. The values of $\rhob$ at the positions of the three systems are estimated assuming a Navarro-Frank-White profile for the MW halo. 
All three are located within $\sim 1.7$ kpc from the Sun, resulting in background densities not dissimilar from the local one, $\rhob \sim 0.3 - 0.4 \,\mathrm{GeV}\,\mathrm{cm}^{-3}$. The extension of the spike varies only mildly across systems and values of $\gamma$ and is of the order $\rsp \sim \mathcal{O} \left( 1 \,\mathrm{pc} \right)$.
In summary, we model the distribution of dark matter around each BH as
\begin{align}
\label{eq:spikes}
  \rhoDM \left(r \right) = 
    \begin{cases}
      \rhosp(r) \,,  & r \leq  \rsp  \\
      \rhob  \,, & r \geq \rsp \,, 
    \end{cases}
\end{align}
where $\rhosp$ is the spike profile
\begin{equation}
    \rhosp(r)=\rhob \left( \dfrac{r}{\rsp} \right)^{-\gamma} \, . 
\end{equation}
In practice, to reduce computational cost, in our simulations we define a truncation radius $r_\mathrm{t} = 20 \, r_\mathrm{orb}$, beyond which the slope transitions from $\gamma$ to $\gamma +2$. This allows us to simulate only the inner regions of the spike which are relevant for the evolution of the binary (the profile transition begins at distances from $r_\mathrm{orb}$ well beyond the estimated maximum impact parameter~\cite{Kavanagh:2024lgq,Kavanagh:2020cfn}).

\begin{figure*}[th]
\centering
\includegraphics[width=1\linewidth]{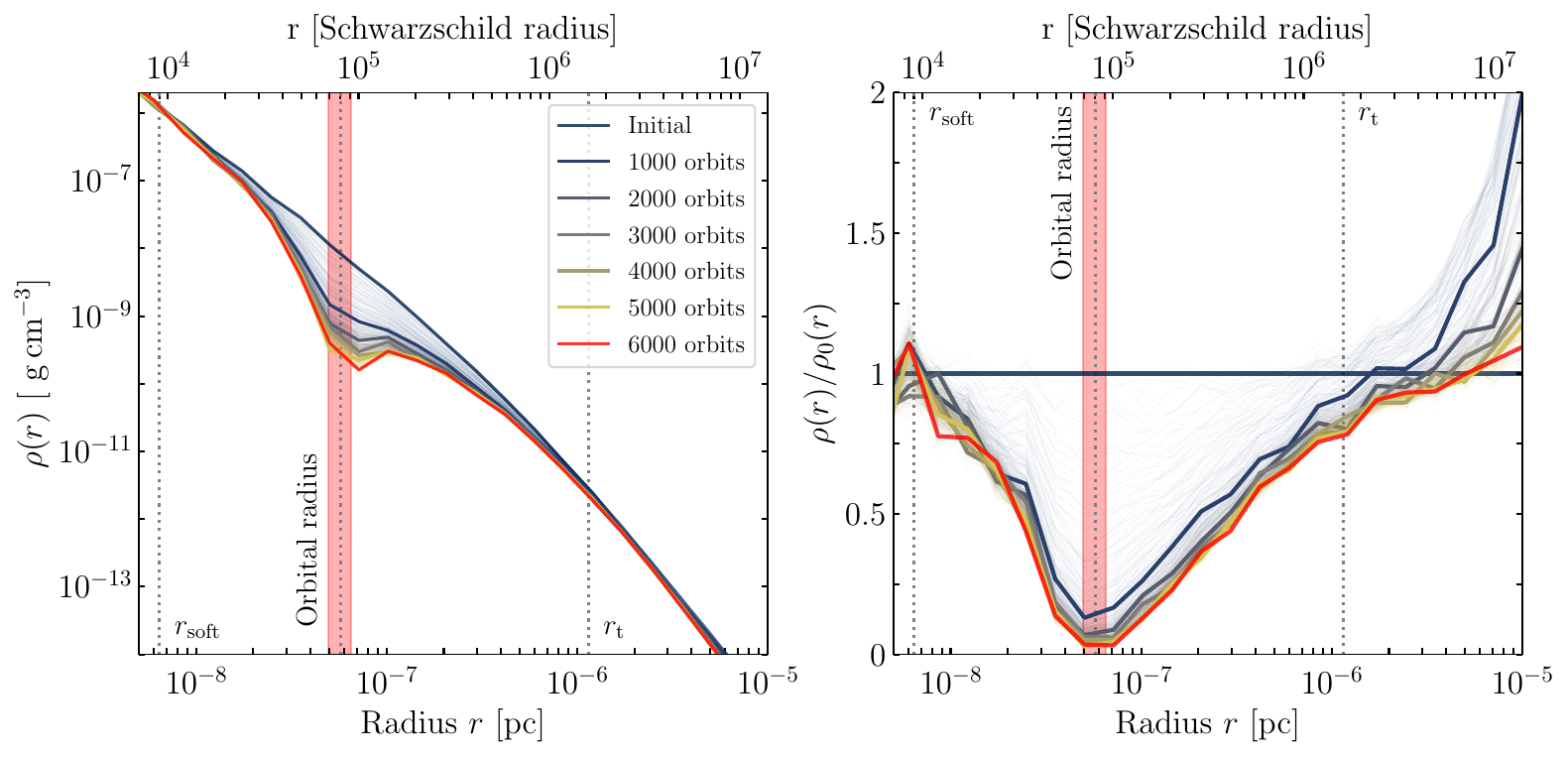}
\caption{\textbf{Time evolution of the DM density profile} around the BH, for one realisation of system A with initial density profile index $\gamma=9/4$. The red band around the orbital radius indicates the size of the star. Dotted lines indicate the BH softening scale $r_\mathrm{soft}$ and the truncation radius $r_\mathrm{t}$. Thin lines are ten orbits apart, the simulation spans 6000 orbits. \textbf{Left}: DM density as a function of radius. \textbf{Right}: ratio of the final and initial density profiles. Notice that the variation of the orbital radius over 6000 orbits is much too small to be appreciated on this figure.}
\label{fig:depletion}
\end{figure*}

\subsection{Simulating with \textsc{NbodyIMRI}}
\label{sec:Setup}

The code \textsc{NbodyIMRI}~\cite{NbodyIMRI,Kavanagh:2024lgq} is an $N$-body simulator conceived precisely to study the gravitational evolution of a binary system embedded in a background of dark matter particles. 
The code follows the Newtonian evolution of the primary and secondary mass of the binary system and of each of the DM pseudo-particles forming the spike. 
At each time step, the velocities and positions of all masses are updated through a fourth-order leapfrog algorithm (see Ref.~\cite{Kavanagh:2024lgq} for details).
The code computes the accelerations taking into account the gravitational forces generated: (a) by the two orbiting bodies (star and central BH, in this case) on each other (b) by each DM particle on the secondary mass (star) (c) by the two main orbiting bodies on the DM particles. Pairwise interactions between DM particles are neglected, and so is the force generated by the DM spike on the central body. These approximations are justified by the fact that the total DM mass inside the orbital radius (typically $\ll 10^{-6}\,M_\odot$) is negligible compared with the masses of the primary and secondary. Gravitational wave emission from the binary is not included in the simulations. 
DM particles are treated as point masses; the potentials of the BH and of the star are instead softened by distributing their mass within spheres of uniform density. In the case of the latter, the natural softening radius is set by the stellar radius $R_\star$ (Table~\ref{tab:observations}); for the former, we introduce an artificial softening scale $r_\mathrm{soft}$.

By taking into account the forces generated on the orbiting star by all DM particles, the simulator automatically captures the drag described by dynamical friction. 
Furthermore, the simulation follows the acceleration of the DM particles induced by interactions with the star and their displacement from the orbital region. It is therefore able to consistently account for the depletion of the spike and the corresponding reduction of the drag on the star, what we refer to as the feedback effect.

We populate the DM spikes extracting the initial positions and velocities of 20k particles (or 25k for the highest $\gamma$ value, see \cref{sec:app:simulations}).  
The number of particles, in combination with the total spike mass, determines the resolution of the simulation. Since we keep the number of particles fixed and the total mass increases with $\gamma$,  low-$\gamma$ spikes are better resolved. Velocities are initialised self-consistently through the Eddington inversion method (see Ref.~\cite{Kavanagh:2024lgq} for a complete discussion of the initialisation procedure). 
The masses $M_\star$ and $\MBH$ are set to the central values in reported Table~\ref{tab:observations}. The star is initialised on a circular orbit; its orbital speed is obtained from the observed period and radius, which are set to the central values  in Table~\ref{tab:observations}. 

\begin{figure*}[tb]
\centering
\includegraphics[width=0.98\linewidth]{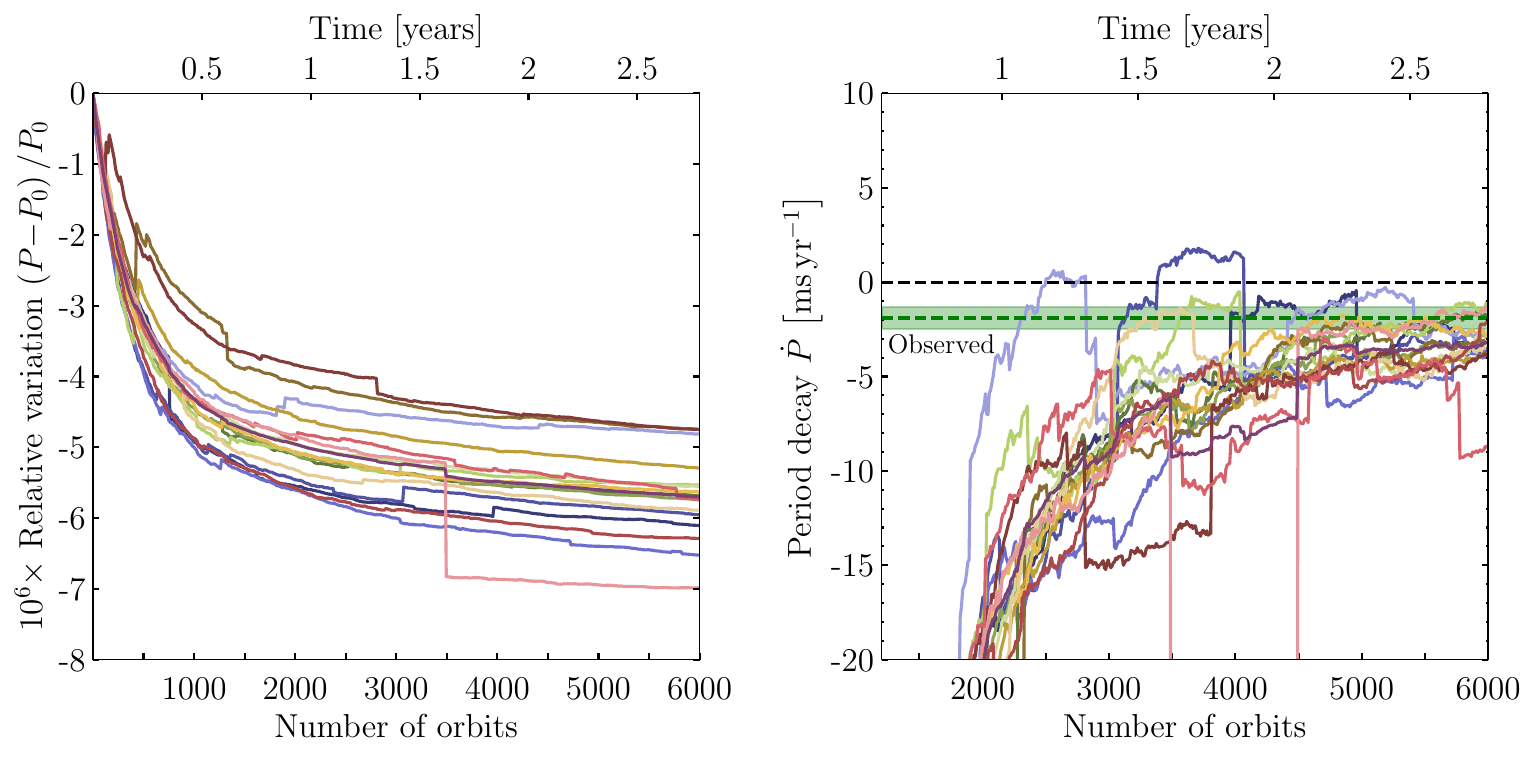}
\caption{\textbf{Evolution of the orbital period} for 16 realisations of System A, assuming a spike index of $\gamma=9/4$ (same as \cref{fig:depletion}). \textbf{Left}: fractional change of the orbital period $P$ with respect to the initial value $P_0$, as a function of time.  \textbf{Right}: period decay rate $\dot{P}$ as a function of time, averaged over 1000 orbits to reduce noise. The final points of each line (6000 orbits) correspond to our estimates of the period decay rate $\dot{P}_\gamma$.}
\label{fig:decay}
\end{figure*}

We vary the spike index $\gamma$ between 1.9 and 2.45, with a step of 0.05\footnote{In the case of Systems B and C, this range includes a few values of $\gamma$ already ruled out by the estimate depicted in Fig.~\ref{fig:disruption_timescale}.}. For each value of $\gamma$, we generate 16 different realisations of the spike. We follow the evolution of each realisation for 6000 orbits. 
The positions and velocity of the BH and star are saved at fixed time step intervals every 10 orbits, together with the radially binned density distribution of DM particles.
More technical details on the simulations are given in \cref{sec:app:simulations}.

\subsection{Results}
\label{sec:Results}
\subsubsection{Dark matter depletion}

In~\cref{fig:depletion} we show the typical evolution of a DM spike density profile. The profiles shown are snapshots obtained from a single simulation of System A, with the spike index set to $\gamma =9/4$. The vertical band is centred at the orbital radius and its width corresponds to the size of the star $R_\star$. As the simulation proceeds, the spike density is progressively depleted around this region.
Depletion progresses rapidly at the onset, with most of it occurring during the first few hundred orbits. This quick variation is followed by a more stable phase: after a few thousand orbits, the DM density settles to a value around two orders of magnitude lower than the initial one.
We observe a similar behaviour in all simulations. The total amount and rapidity of the depletion presents some variability across the three systems.
As expected, we observe the highest depletion (and fastest stabilisation) in System C, which is characterised by the highest mass ratio $q$. The density in this case is depleted by almost three orders of magnitude by the end of the run.

We note that the simulations span a period of a few years only. The fast initial depletion phase we see in \cref{fig:depletion} occurs over a time scale of $\sim\mathcal{O}(1\mathrm{yr})$. This scale is much shorter than the lifetimes of the systems, or than the time scales over which the orbital radius is expected to vary appreciably (we estimate $\sim \mathcal{O}(10^{5}-10^{6} \mathrm{yr})$ for a $1\%$ variation at the reported decay rates). 
This separation of timescales implies that the depletion process reaches a quasi-stationary configuration effectively instantaneously relative to the binary's orbital evolution.
Thus in the physical systems, this fast depletion phase is not dynamically observable.

\subsubsection{Period decay rate}

Next, we extract from the simulation data the variation of the orbital period $\dot{P}$. 
For each snapshot, we compute the orbital parameters from the relative velocities and positions of the BH and star. The orbit of the star is found to remain circular to high accuracy, while the semimajor axis decreases in time, as expected. From the orbital parameters, we calculate the orbital period $P$. 
In the left panel of~\cref{fig:decay}, we show the fractional change of $P$ with time, for 16 realisations of System A with $\gamma=9/4$. As expected, at first the period decays sharply and  non-linearly, corresponding to the phase of fast initial depletion of the spike. As the spike density stabilises, the system settles to a more steady decay rate. Discrete jumps in the period are due to rare close encounters between DM pseudo-particles and the orbiting star.

To extract the rates of orbital period decay $\dot{P}$, we compute the ratio $(P_{i+1} - P_i )/ \Delta t$ , subtracting the values obtained from successive snapshots $P_i$ and dividing by the corresponding time interval. We average this ratio over 1000 orbits, in order to reduce noise. 
The right panel of~\cref{fig:decay} depicts the estimated $\dot{P}$ as a function of time, for the same set of simulations depicted in the left panel. As before, each curve corresponds to a different realisation of the system, while the horizontal band indicates the observational constraint on the period decay. 
The smoothing procedure does alter the value of $P$ and $\dot{P}$ significantly over the first few thousand orbits. However, we are not interested in the initial (non-physical) transient phase, but in the more stable configuration reached at the end of the simulation. We have checked that varying the smoothing window from 1000 to 100 orbits has a negligible effect on the final $\dot{P}$ readings.

\begin{figure}[!ht]
\centering
\includegraphics[width=\linewidth]
{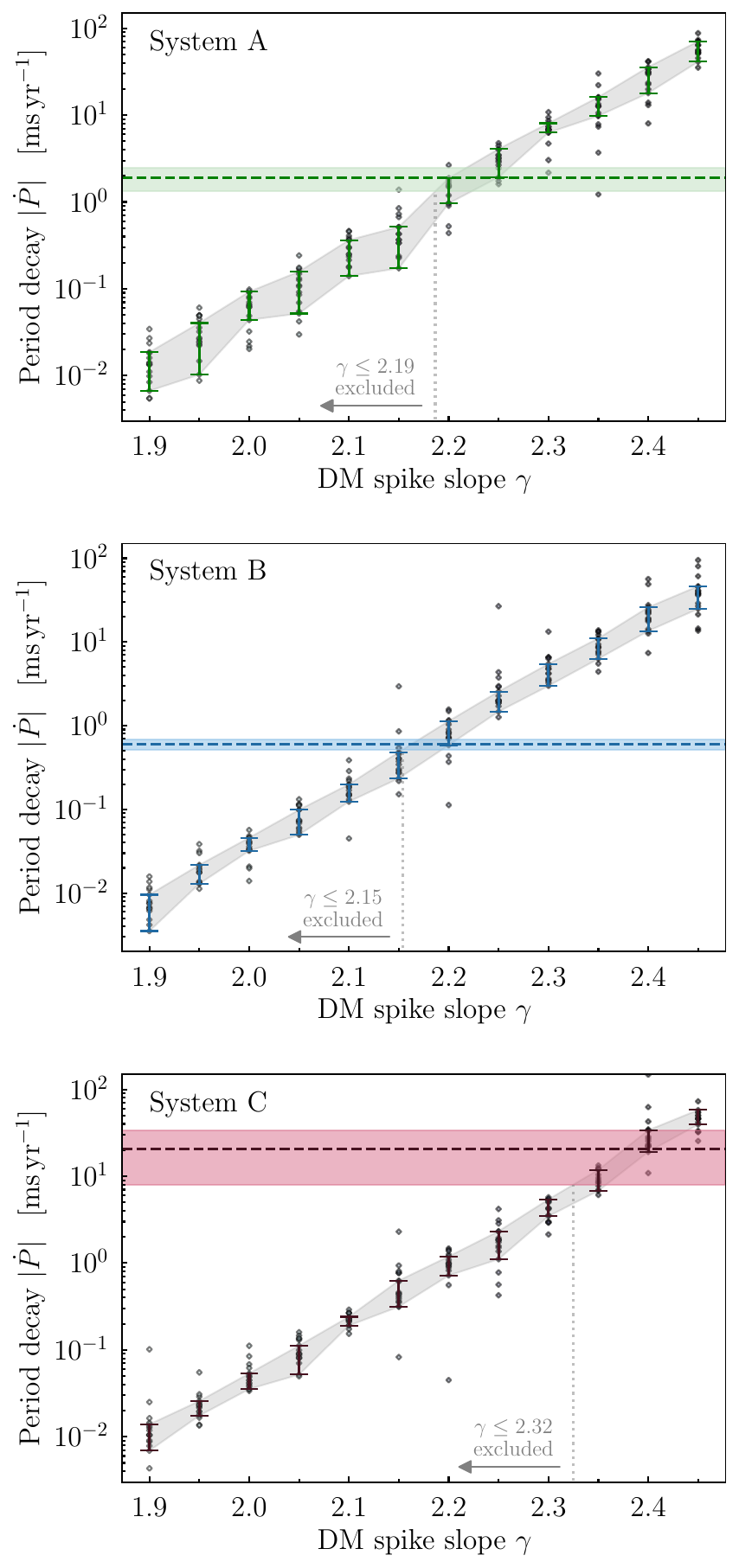}
\caption{\textbf{Period decay rates extracted from simulations} as  a function of the spike index $\gamma$. Diamonds mark the decay rates inferred from single realisations, while the vertical error bars span the interval covered by the 11 most central ones, corresponding to $\approx 68\%$ of points. In each panel, the horizontal band shows the observational $1\sigma$ constraint. Values of the spike index that predict a period decay rate lower than the observed one are unable to explain the anomalous decay rates. Arrows indicate the $\gamma$ range that is ruled out by this criterium. 
}
\label{fig:Pdot}
\end{figure}
We take the value of $\dot{P}$ attained at the end of the simulation to approximate the physical period decay of the systems, and refer to this quantity as $\dot{P}_\gamma$. 

For a given value of $\gamma$, we obtain 16 independent estimates of $\dot{P}_\gamma$ from the 16 realisations
(in the right panel Fig.~\ref{fig:decay}, these are the end points of each curve). 
The spread between points is due to the intrinsic discretisation noise of the simulations. In order to quantify this error, we select the 11 elements out of 16 which are closest to the median. The upper and lower extremes of this set
define a $\sim68\%$ confidence level (CL) interval. 
We choose to define the CL interval this way, rather than compute the mean value of $\dot{P}_\gamma$ and dispersion of the measurements. Our choice provides a more robust estimator for the confidence interval, as it is not subject to fluctuations induced by outliers (simulations in which the orbit of the star deviated strongly due to a single close encounter with one of the DM particles). 

In~\cref{fig:Pdot} we show the final period decay rates $\dot{P}_\gamma$, each panel corresponding to one LMXB system. For each value of $\gamma$, results of the single simulations are shown as points. The superposed vertical bars indicate the $68\%$ CL intervals. 
One-sigma observational constraints on the decay rates are indicated as horizontal bands. 
We see from the simulations that the magnitude of $\dot{P}_\gamma$ increases with $\gamma$. This is explained almost entirely by the variation of $\rhoDM(r_\mathrm{orb})$ between simulations with different $\gamma$. This suggests that that the variation with $\gamma$ of the velocity distribution, and therefore of $\xi(v)$ in Eq.~\eqref{eq:Pdot}, have little effect on the orbital decay. 

For each of the three systems, we can visually identify a few values of $\gamma$ for which the simulations are in best agreement with the experimental measurement. In the case of System A, simulations with $\gamma =2.2$ and $\gamma =2.25$ give the best fits; System B prefers slightly lower values, $\gamma =2.20$ being the only value for which the $68\%$ CL error bar overlaps with observations. System C, whose decay rate is larger, requires larger spike indexes than the other two systems, $\gamma =2.35$ and $\gamma =2.4$.

\subsubsection{Constraints on the spike profile}

Spikes for which $|\dot{P}_\gamma|$ is smaller than the observed rate cannot explain the anomalous decays. Feedback happening over the simulation time is sufficient to prevent these spikes from sustaining the dynamical friction required to explain the anomalous decays.
To obtain a lower limit on $\gamma$, we construct the $68\%$ CL confidence band by connecting the upper and lower extremes of the error bars with a piecewise linear function. The lower-left 
intersection of the simulated and observed confidence bands provides a $\sim$1$\sigma$ lower limit on $\gamma$.
This analysis allows us to rule out spike indexes up to  $\gamma \lesssim 2.15-2.20$ for Systems A and B and $\gamma \lesssim 2.3$ for System C as responsible for the anomalous decay rates. 
Note that the confidence intervals reported here account for the limited resolution of the simulations, our limited statistics and the inherent stochasticity of the process. However, there are also (correlated) observational uncertainties on the component masses and orbital radius (see Table~\ref{tab:observations}) at the level of a few percent (or larger in the case of System C), which we do not account for here. 
We could in principle obtain an upper limit on the required spike index $\gamma$ with the same method we use to compute the lower limit. However, we cannot exclude that further evolution of the systems will lead to larger spike depletion and reduce decay rates further (see \cref{sec:Discussion}).

\begin{figure*}[tb]
\centering
\includegraphics[width=0.8\linewidth]{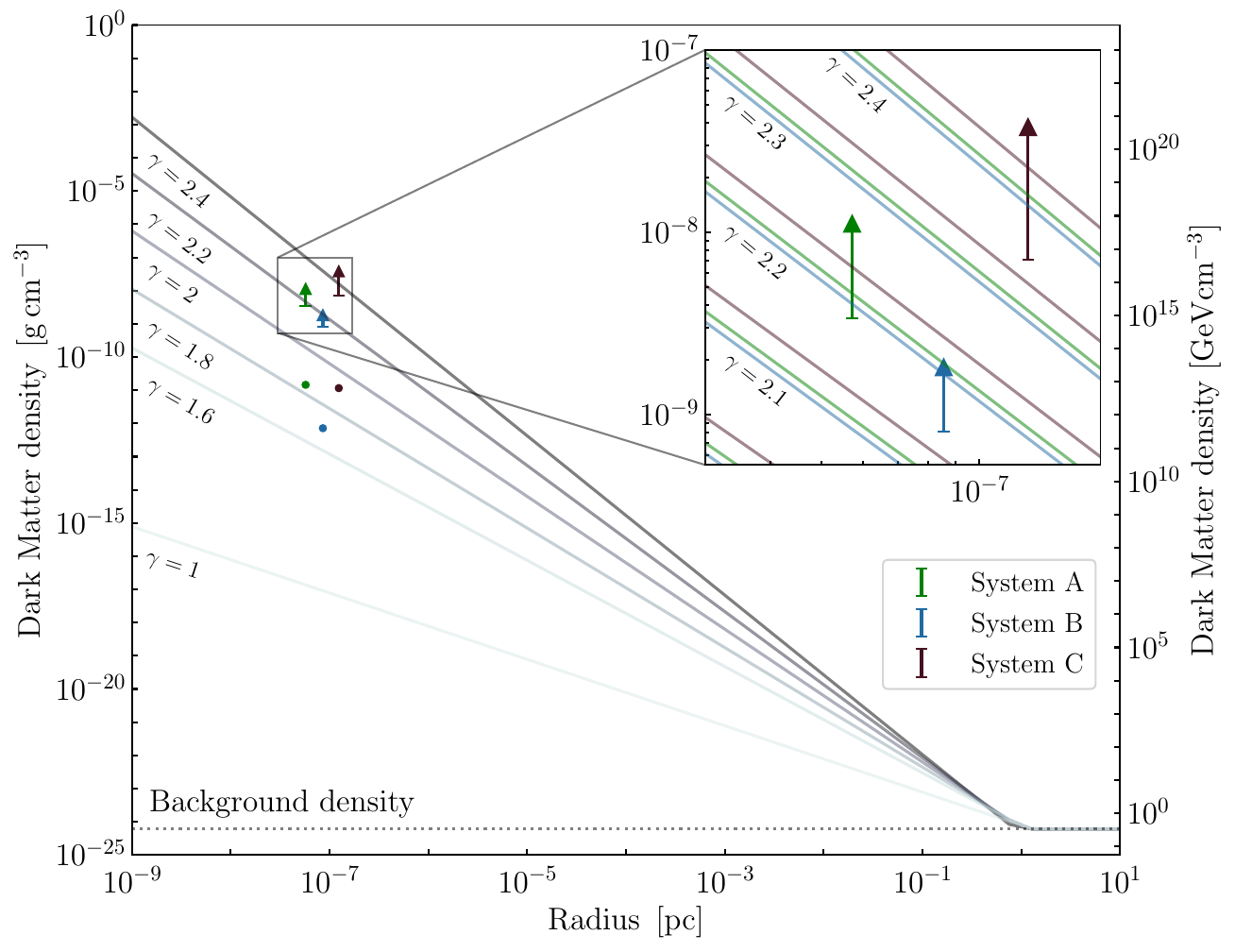}
\caption{\textbf{Limits on dark matter density profiles} compatible with the observed decay rates of Systems A, B and C. The DM density predicted by Eq.~\eqref{eq:spikes} is shown as a function of the distance from the BH, for a set of values of the power-law index $\gamma$. The three points indicate, for each system, the value of the density at the orbital radius inferred through~\cref{eq:Pdot}, neglecting the effects of feedback. The green and blue dots, obtained for Systems A and B  respectively, are consistent with the density values inferred by Ref.~\cite{Chan:2022gqd}. The coloured bars indicate the lower limits obtained from simulations, accounting for the depletion of the DM spike by the star. Simulations on longer timescales may lead to further depletion and potentially push these limits to larger DM densities. 
To avoid overcrowding, in the main panel the density profiles are shown only for System A. The small differences between profiles for the three systems can be appreciated in the inset, where they are shown in different colours.}
\label{fig:profiles}
\end{figure*}

In \cref{fig:profiles} we report the constraints on the spike density at the orbital radii of the three stars.
The similarity of the three systems is such that the density profiles $\rho(r)$ vary little across systems. In the main figure we only show the profiles corresponding to System A for visual clarity (the difference between profiles for the three systems can be appreciated in the inset). 
The coloured points show the density necessary to generate the observed $\dot{P}$, as obtained using the Chandrasekhar expression \cref{eq:Pdot}, without taking into account feedback. For comparison with previous works~\cite{Chan:2022gqd,Kar:2023vqe,Ireland:2024lye}, we set $\Lambda = \sqrt{M_\mathrm{BH}/M_\star}$ and $\xi(v)=1$.\footnote{Note that in principle $\xi(v)$ can be calculated explicitly from the spike distribution function~\cite{Karydas:2024fcn} and that more recent simulations are better fit by $\Lambda \approx 0.3 M_\mathrm{BH}/M_\star$~\cite{Kavanagh:2024lgq}.} Points corresponding to systems A and B are in agreement with the values reported by Chan \& Lee~\cite{Chan:2022gqd}. The density required for system C is computed here for the first time. 
Interestingly, despite the much larger decay rate observed, the DM density required for system C is close to the one required by system A.

We report the lower limits obtained from our simulations as vertical arrows. 
If interpreted as measurements of the DM density profile, the three observations give strikingly compatible results. 
However, we underline again that these results should not be regarded as confidence intervals, but rather as lower limits to the initial spike density required to sustain the observed $\dot{P}$.
These lower limits are about three orders of magnitude larger than the values obtained from the naive Chandrasekhar analysis, showing the crucial role played by feedback effects in this context. The largest shift occurs for System C, followed by System B and System A. This is consistent with the expectation that larger mass-ratios are associated to higher depletion and feedback effects. 

\section{Discussion}
\label{sec:Discussion}

\begin{figure}[tb]
\centering
\includegraphics[width=0.98\linewidth]{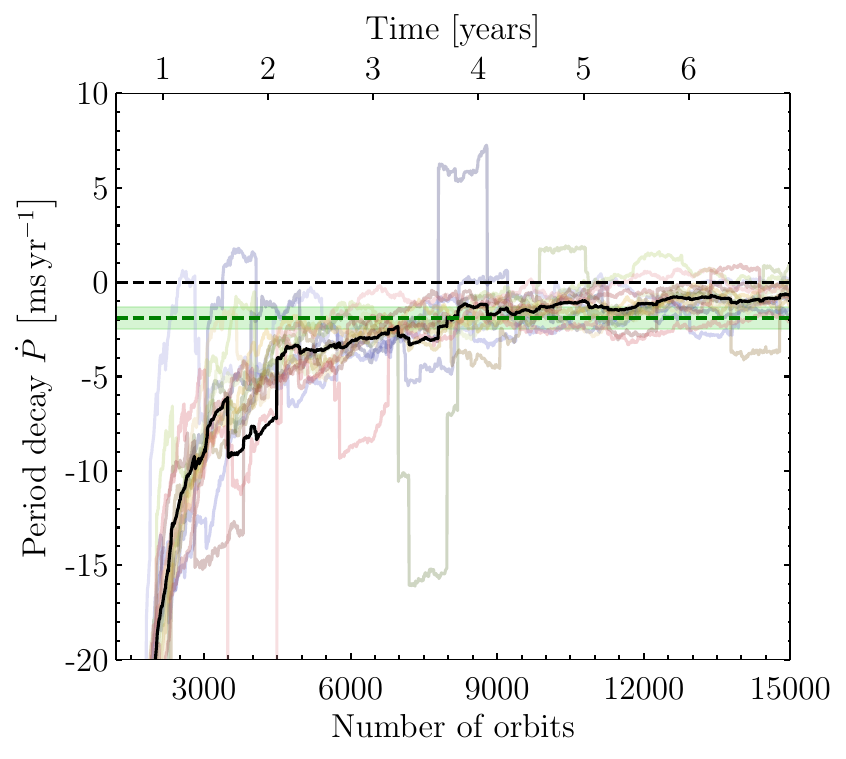}
\caption{\textbf{Evolution over 15k orbits} of 16 realisations of System A, for $\gamma=2.25$. The figure shows the variation $\dot{P}$ of the orbital period as a function of time. The black curve is the average taken across the 16 realisations, while individual readings are shown as light coloured curves. The values of $|\dot{P}|$ read at 15000 orbits are lower than those read at 6000 orbits by a factor of a few.}
\label{fig:long_decay}
\end{figure}

In the previous section, we analysed simulations of binary systems performed over $6000$ orbits, covering physical timescales of a few years, shorter than the observations of these systems (which typically cover tens of thousands of orbits over tens of years). In fact, these timescales are also much shorter than age of the binary systems under consideration. For example, the metallicity of the secondary star in System A suggests that the system was formed in the Galactic disk before being ejected by the natal kick of the BH~\cite{Mirabel:2001ay,2006ApJ...644L..49G}. Stellar evolution modelling suggests that this system was formed $2-5 \,\mathrm{Gyr}$ ago~\cite{Gualandris:2004tv}, while the kinematics of the system suggest that it last passed through the Galactic Disk at least 11 Myr ago (providing an independent bound on the age of the system). In addition, assuming a constant decay rate we can estimate the time scale for the orbital radius to vary, $\Delta t \approx  (3P/2\dot{P}) \times (\Delta r_\mathrm{orb} /r_\mathrm{orb})$, corresponding to $\Delta t \approx \mathcal{O}(10^5-10^6)\,  \mathrm{yr}$ for a $\sim 1\%$ variation.
These timescales are many orders of magnitude larger than the duration of our simulations. Therefore, in order to interpret the output of our simulations as a confidence interval on the value of the initial spike density, it would be necessary to prove that the simulated system has reached a stationary or quasi-stationary configuration.

To study the further time evolution of the spike, we have run simulations of the three systems for a larger number of orbits (15000), only for the value of $\gamma$ providing the best fit to observations at the end of the 6000-orbit runs. 
The results of the long run for System A are shown in \cref{fig:long_decay}. The $\dot{P}$ value extracted at the end of the long run is a factor of a few smaller than the one we report in our results, reflecting the fact that the spike is slowly being depleted further. Because of the strong scaling of the DM density with $\gamma$, this factor of a few change in $\dot{P}$ would translate to a shift in the inferred value of $\gamma$ of $\Delta \gamma \sim 0.05$. While the rate of change of $\dot{P}$ is decreasing with increasing simulation time,  we have not been able to verify that $\dot{P}$ has reached a stationary value. Therefore, we report our results only as lower limits on the initial DM density. The long term evolution of the spikes will be discussed in a separate work (in preparation).

So far, we have studied the possibility that the anomalous orbital decay of these LMXBs may be due to large overdensities of DM. However, we note that DM spikes are not generically expected in such systems. DM spikes around astrophysical BHs are often assumed to form due to adiabatic growth, in which light seed BHs grow slowly by many orders of magnitude at the centre of a DM halo~\cite{Gondolo:1999ef,Bertone:2024wbn}. While such a mechanism could be invoked for SMBHs at the centres of galaxies, it does not predict the presence of DM spikes around stellar mass BHs such as those in the systems we study here. In addition, neither over-dense DM `cusps' formed at the centre of DM halos~\cite{Delos:2019mxl,Delos:2022yhn} nor ultra-compact minihalos (formed from large primordial density perturbations)~\cite{Bringmann:2011ut,Delos:2017thv} are expected to be associated with such light BHs. 

Also unclear is the possible formation mechanism of LMXBs containing BHs as the primary~\cite{Yungelson:2006dn}. The formation of DM spikes typically requires symmetric and quiescent environments (see e.g.~\cite{Ullio:2001fb,Bertone:2005hw}). This suggests that for these systems to host a DM spike, the DM and its spike would have to form in isolation (through some as-yet-unknown mechanism in the late Universe), followed by the dynamical capture of the star. However, tidal capture is not expected to be effective in the absence of nearby dense stellar clusters~\cite{Li:2015kra} as is the case for the systems we study here. 

Instead, the standard formation scenario for BH-LMXBs begins with a binary of main sequence stars, the more massive of which eventually expands as it evolves off the main sequence. The secondary may then be engulfed by the expanding primary, with friction from this `common envelope' shrinking the orbit and, potentially, ejecting this stellar envelope~\cite{2006csxs.book..623T,Li:2015kra}. The core of the primary may then proceed to collapse to a BH and if the binary survives this process it can continue to develop into an LMXB. To date, no mechanisms have been proposed for the formation of DM spikes around stellar binaries and it is hard to see how a cold, dense DM spike could survive the violent processes involved in this standard picture of BH-LMXB formation. 

An alternative possibility is that the primaries in these systems are stellar-mass \textit{primordial} black holes (PBHs), which formed long before matter-radiation equality. These are generically expected to host dense DM spikes with a $\gamma = 2.25$ density profile~\cite{Mack:2006gz,Ricotti:2007jk,Adamek:2019gns}. It has also been recently proposed that friction from the DM spike can lead to dynamical capture of stars, leading to the formation of PBH LMXRBs in the Milky Way~\cite{Kritos:2020wcl,Esser:2025kua}.
The density of DM spikes around PBHs is fixed by the dynamics of DM in the early Universe and is expected to be $\mathcal{O} (10^{-6})\,\mathrm{g}/\mathrm{cm}^{3}$ at the orbital radius of these systems. This density is larger than any of the cases we have considered so far and from Fig.~\ref{fig:profiles} it is clear that this scenario cannot be excluded based on the simulations we have presented here. 
Reference~\cite{Ireland:2024lye} explored a PBH-and-spike scenario for these systems, using the semi-analytic \texttt{HaloFeedback} formalism to account for feedback~\cite{Kavanagh:2020cfn}. 
We note that in $N$-body simulations the timescale for feedback effects is typically longer than in the \texttt{HaloFeedback} formalism~\cite{Mukherjee:2023lzn,Kavanagh:2024lgq}. Nevertheless Ref.~\cite{Ireland:2024lye} invoked DM kinetic decoupling in order to further suppress the spike density and match the observed period decay rates.
In summary, with our simulations we cannot definitively determine whether the anomalous decay of these systems can be explained by a PBH-and-spike scenario. A definitive conclusion requires the extrapolation of feedback effects to longer time scales and is left for future work.

As mentioned in \cref{sec:Anomalies}, purely astrophysical mechanisms have been proposed to explain the observations. The circumbinary‑disk (CB) scenario posits a gaseous disk outside the binary that exchanges angular momentum with the inner binary through resonant torques; observations have reported possible CB signatures in Systems A and B~\cite{2014ApJ...788..184W}. 
CB torques combined with anomalous magnetic braking can reproduce the high $\dot{P}$ for Systems A and B, although they produce donor effective temperatures higher than the spectral types indicate~\cite{ 2019ApJ...876L..11C}.  
An alternative avenue is provided by the convection and rotation boosted (CARB) prescription for MB, a recent, physically motivated revision of the standard prescription that amplifies angular‑momentum loss.
Evolution studies using CARB match System A's period decay rate, but significantly overestimate $\dot{P}$ for System B and are unable to match the decay rate of System C~\cite{2024ApJ...976..210F}.

In summary, purely astrophysical descriptions are at present unable to consistently account for all of the observations.  Meanwhile, the DM spike scenario cannot be entirely ruled out based on the results of our simulations. Indeed, the minimum slopes $\gamma$ required to match the observed period decay rates are strikingly close to one another and to typical values usually associated to DM spikes. The presence of large DM overdensities in these systems would challenge standard spike formation scenarios -- in the case of astrophysical BHs -- and have important implications for the field of DM. However, as we discuss above, we caution the reader that further study of the long-term evolution of these systems would be necessary in order to claim evidence of DM spikes.

\section{Conclusions}
\label{sec:Conclusions}

We have critically assessed the claim that anomalous period decay rates observed in BH-LMXBs can be explained by dynamical friction caused by dark matter (DM) spikes surrounding the black holes (BHs). Crucially, the motion of the binary leads to feedback in the DM spike. Analytic estimates suggest that shallow DM spikes, with slopes $\gamma \lesssim 1.75 - 2.00$, would be disrupted on timescales shorter than the duration of observations of LMXBs and thus cannot explain the anomalous period decay rates. 

Motivated by this, we have performed simulations of the three binaries listed in Table~\ref{tab:observations} -- System A (XTE J1118+480), System B (A0620-00) and System C (NM1991) -- embedded in dark matter spikes, varying the slope of the density profile. Simulations have allowed us to quantify feedback effects, which reduce the dark matter density and the magnitude of the dynamical friction force exerted on the orbiting star. Based on these results, we conclude that spikes with slopes lower than $\gamma \lesssim 2.2$ for Systems A and B and $\gamma \lesssim 2.4$ for System C are not able to sustain the friction required to cause the reported anomalous period decays (see Fig.~\ref{fig:profiles}).
Drawing conclusions over steeper DM profiles requires extrapolating feedback effects to time scales longer than can be currently simulated. This extrapolation will be discussed in a forthcoming work. 

\subsection*{Acknowledgements}

The authors acknowledge funding from the \textit{Consolidaci\'on Investigadora} Project \textsc{DarkSpikesGW}, reference CNS2023-144071, financed by MCIN/AEI/10.13039/501100011033 and by the European Union ``NextGenerationEU"/PRTR.
The authors also acknowledge Santander Supercomputing support group at the University of Cantabria who provided access to the supercomputer Altamira at the Institute of Physics of Cantabria (IFCA-CSIC), member of the Spanish Supercomputing Network, for performing simulations.
BJK also acknowledges support from the project SA101P24 (Junta de Castilla y León).
The authors thank Gianfranco Bertone for helpful discussions on this research.

\bibliography{main}

\appendix

\section{Simulation setup}
\label{sec:app:simulations}
In this section we provide additional details about the simulation settings. 

\subsection{Softening} To prevent the occurrence of very strong scattering events that go beyond the time resolution of the simulations, we soften the gravitational forces generated by the BH on the DM particles. We replace the BH with a uniform distribution of mass over a sphere of radius $r_\mathrm{soft}=3 \times 10^3 r_\mathrm{isco}$. The softening radius is chosen to ensure the stability of the inner spike at the time step we use. As shown in \cref{fig:depletion}, $r_\mathrm{soft}$ is about an order of magnitude smaller than the orbital radius, a region not affected by the depletion caused by the star.
We use the same recipe to describe the gravitational force exerted by the star on the DM particles, distributing its total mass uniformly over the stellar radius. We have verified that the use of different softening prescriptions for the stellar potential does not affect our results.

\subsection{Resolution}

We generate $N_\mathrm{DM}=20000$ DM particles for each simulation, with the exception of the largest value $\gamma=2.45$, for which we set $N_\mathrm{DM}=25000$. The DM particles are distributed randomly following the DM spike distribution profile and each carries an equal fraction of the total spike mass. We softly truncate the spike profile at a radius $r_\mathrm{t} = 20 \times r_\mathrm{orb}$. 
Since the total spike mass grows with $\gamma$ (by about a factor two at each step), the mass of the DM particles also increases. Simulations in which the slope is mildest, i.e. $\gamma=1.90$, are best resolved, with DM particle masses of $M_\mathrm{DM} \sim \mathcal{O} (10^{-12} \Msun)$. The DM particle mass grows with the spike index up to $M_\mathrm{DM} \sim \mathcal{O} (10^{-9} \Msun)$ for the highest values of $\gamma$. The resolution difference across spikes is reflected in the increasingly large error bars we obtain, as shown in \cref{fig:Pdot}. Reducing all errors to the level achieved for lower $\gamma$ values is unfeasible, requiring increasing $N_\mathrm{DM}$ by about three orders of magnitude (run time is linear in $N_\mathrm{DM}$). 
We choose to keep $N_\mathrm{DM}$ fixed to avoid unnecessarily degrading the low $\gamma$ simulations.

\subsection{Time step}

In \textsc{NbodyIMRI}~\cite{NbodyIMRI,Kavanagh:2024lgq}, the time step is set as a fixed fraction of the binary orbital period; we found $\sim$ 10000 steps per orbit to be the minimum requirement to maintain the stability of the systems studied here. 
To reduce CPU time, we implemented hierarchical time-stepping~\cite{2011EPJP..126...55D}:  particles nearer to the central mass are advanced with multiple finer steps per main step, while leaving more distant particles on the base step. At each main step the code selects particles by their distance from the central mass using a tunable threshold $r_\mathrm{in}$. During sub-steps the particles in the inner population are advanced with the same leapfrog integrator as the main code, while the two primary bodies and all outer particles are held fixed. Forces on the outer particles and on the two primaries are evaluated once per main step. The main step and the number of sub-steps per main step are both tunable; in the configuration that produced our results the main step is set to 1000 steps per orbit and split into 10 sub-steps. Inner particles effectively receive the equivalent of 10000 steps per orbit while outer particles advance at the coarser main-step rate. We set the threshold radius $r_\mathrm{in}$ to be 10 times the BH softening radius $r_\mathrm{soft}$. 
We verified stability by evolving the central mass plus DM particles for 3000 orbits and confirming the density distribution remains stable. Overall, we found that this hierarchical time-stepping reduced the total runtime by approximately a factor of five.

\end{document}